\documentclass[amsmath,amssymb,aps,prl,nofootinbib,superscriptaddress,twocolumn,preprintnumbers]{revtex4-2}
\usepackage[utf8]{inputenc}
\usepackage[T1]{fontenc}
\usepackage{mathrsfs}
\usepackage{bm}
\usepackage{url}
\usepackage[normalem]{ulem}
\usepackage{mathtools}
\usepackage{array}
\newcolumntype{P}[1]{>{\centering\arraybackslash}p{#1}}
\newcolumntype{M}[1]{>{\centering\arraybackslash}m{#1}}
\usepackage[caption=false]{subfig}
\usepackage{dcolumn}
\usepackage{graphicx, epsfig}
\usepackage[dvipsnames]{xcolor}
\usepackage{mathrsfs}
\usepackage{bm}
\usepackage{yhmath}
\usepackage[caption=false]{subfig}
\usepackage[normalem]{ulem}
\usepackage{mathtools}
\usepackage{bigints}
\usepackage{float}
\usepackage[colorlinks = true, linkcolor = purple, urlcolor  = blue, citecolor = blue, anchorcolor = blue]{hyperref}
\usepackage{float}
\usepackage{multirow}
\usepackage{tikz,xcolor,hyperref}
\definecolor{darkgreen}{rgb}{0.0, 0.2, 0.13}
\definecolor{bostonuniversityred}{rgb}{0.8, 0.0, 0.0}
\definecolor{lime}{HTML}{A6CE39}
\DeclareRobustCommand{\orcidicon}{
	\begin{tikzpicture}
	\draw[lime, fill=lime] (0,0) 
	circle [radius=0.16] 
	node[white] {{\fontfamily{qag}\selectfont \tiny ID}};
	\draw[white, fill=white] (-0.0625,0.095) 
	circle [radius=0.007];
	\end{tikzpicture}
	\hspace{-2mm}
}
\foreach \x in {A, ..., Z}{\expandafter\xdef\csname orcid\x\endcsname{\noexpand\href{https://orcid.org/\csname orcidauthor\x\endcsname}
			{\noexpand\orcidicon}}
}
\newcommand{\be}{\begin{equation}}
\newcommand{\ee}{\end{equation}}
\newcommand{\ba}{\begin{eqnarray}}
\newcommand{\ea}{\end{eqnarray}}

\def\lp4{$\lambda \phi^4$}
\begin{document}
\preprint{\texttt{FERMILAB-PUB-26-0560-T}}
\title{Constraints on magnetic monopoles from X-ray observations of neutron stars}
\author{Mainak Mukhopadhyay\hspace{-1mm}\orcidA{}}
\email{mainak@fnal.gov}
\affiliation{Astrophysics Theory Department, Theory Division, Fermi National Accelerator Laboratory, Batavia, Illinois 60510, USA}
\affiliation{Kavli Institute for Cosmological Physics, University of Chicago, Chicago, Illinois 60637, USA}
\author{Daniele Perri\hspace{-1mm}\orcidB{}} \email{daniele.perri@fuw.edu.pl}\thanks{Equal contribution.}
\affiliation{Institute of Theoretical Physics, Faculty of Physics, University of Warsaw,
ul. Pasteura 5, PL-02-093 Warsaw, Poland
}
\author{Edward W. Kolb\hspace{-1mm}\orcidC{}}
\email{rocky.kolb@uchicago.edu}
\affiliation{Kavli Institute for Cosmological Physics, University of Chicago, Chicago, Illinois 60637, USA}
\affiliation{Enrico Fermi Institute, University of Chicago, Chicago, Illinois 60637, USA}
\date{\today}
\begin{abstract}
Magnetic monopoles are captured efficiently by neutron stars, and if they catalyze nucleon decay, the decay products would thermalize and generate observable X-ray surface emission. We use archival Chandra, XMM-Newton, and Swift-XRT data for old isolated millisecond pulsars to place conservative limits on the Galactic monopole flux ($F_M$). For a benchmark cross-section of $\sigma_{\Delta \rm B} \sim 10^{-27}\ {\rm cm^2}$, our constraint as a function of monopole mass $m_M$ is given by $F_{\rm M}(m_M) \lesssim  6 \times 10^{-19}~\mathrm{cm^{-2}s^{-1}sr^{-1}} \times \max \big(4 \times 10^{-6}, \min (2 \times 10^{11}~\mathrm{(GeV/c^2)}/m_{\rm M} , 1 ) \big)$. These limits improve previous neutron-star bounds, provide the strongest constraints to date on $F_M$ for $m_M$ between $10^{11} - 10^{13}\ {\rm GeV/c^2}$, and are competitive to existing constraints for this scenario. We also derive complementary constraints from the measured thermal emission of the Magnificent Seven. Our results demonstrate that neutron star X-ray observations provide a powerful probe of magnetic monopoles and motivate dedicated X-ray searches for old neutron stars as a means to test monopole-induced heating.
\end{abstract}
\maketitle
\textit{\textbf{Introduction.}}
Magnetic monopoles have long occupied a central role in particle physics since Dirac first showed that their existence would provide an explanation for electric charge quantization \cite{Dirac:1931kp}. They arise naturally in a wide class of ultraviolet completions of the Standard Model. In particular, ’t Hooft \cite{tHooft:1974kcl} and Polyakov \cite{Polyakov:1974ek} independently demonstrated that spontaneously broken non-Abelian gauge theories admit finite-energy monopole solutions, which appear as topological defects of the vacuum manifold. Such ’t Hooft–Polyakov monopoles are a generic feature of many Grand Unified Theories (GUTs) \cite{Goddard:132994,Dokos:1979vu,Dawson:1982sc} and therefore constitute a well-motivated target for experimental searches.
Across the last several decades, searches for monopoles have been performed spanning from direct searches in experiments~\cite{MACRO:2002jdv,IceCube:2021eye,PierreAuger:2016imq,MoEDAL:2021vix,Super-Kamiokande:2012tld,Perri:2025qpg} to the search for astrophysical and cosmological signatures~\cite{Parker:1970xv,Turner:1982ag,Parker:1987,Adams:1993fj,Kobayashi:2023ryr,Gould:2017zwi,Khelashvili:2026hld}. However, no confirmed detection has been reported to date. Therefore, attempts to narrow and constrain the parameter space continue.

\begin{figure*}[ht!]
\centering
\includegraphics[width=0.49\textwidth]{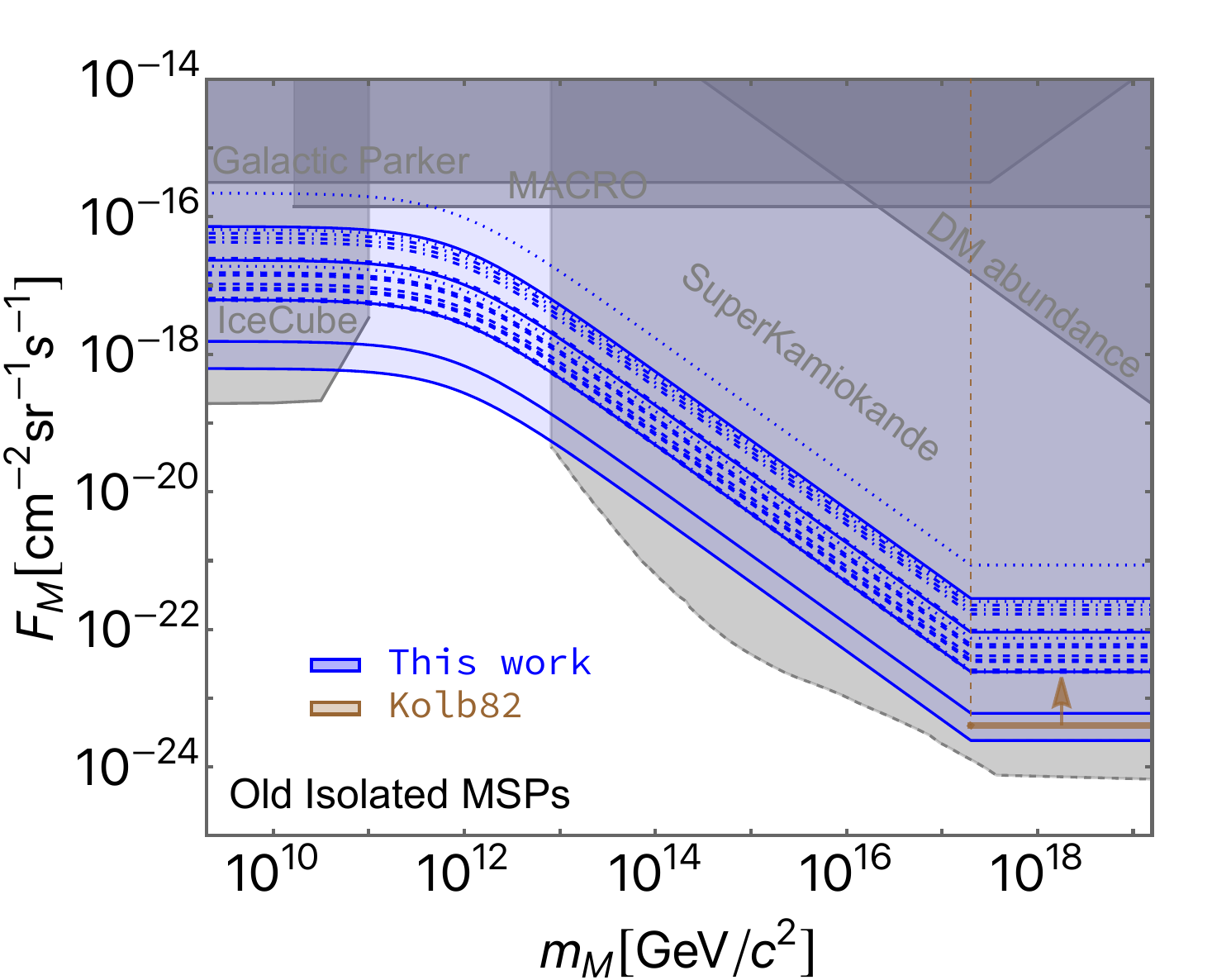}
\includegraphics[width=0.49\textwidth]{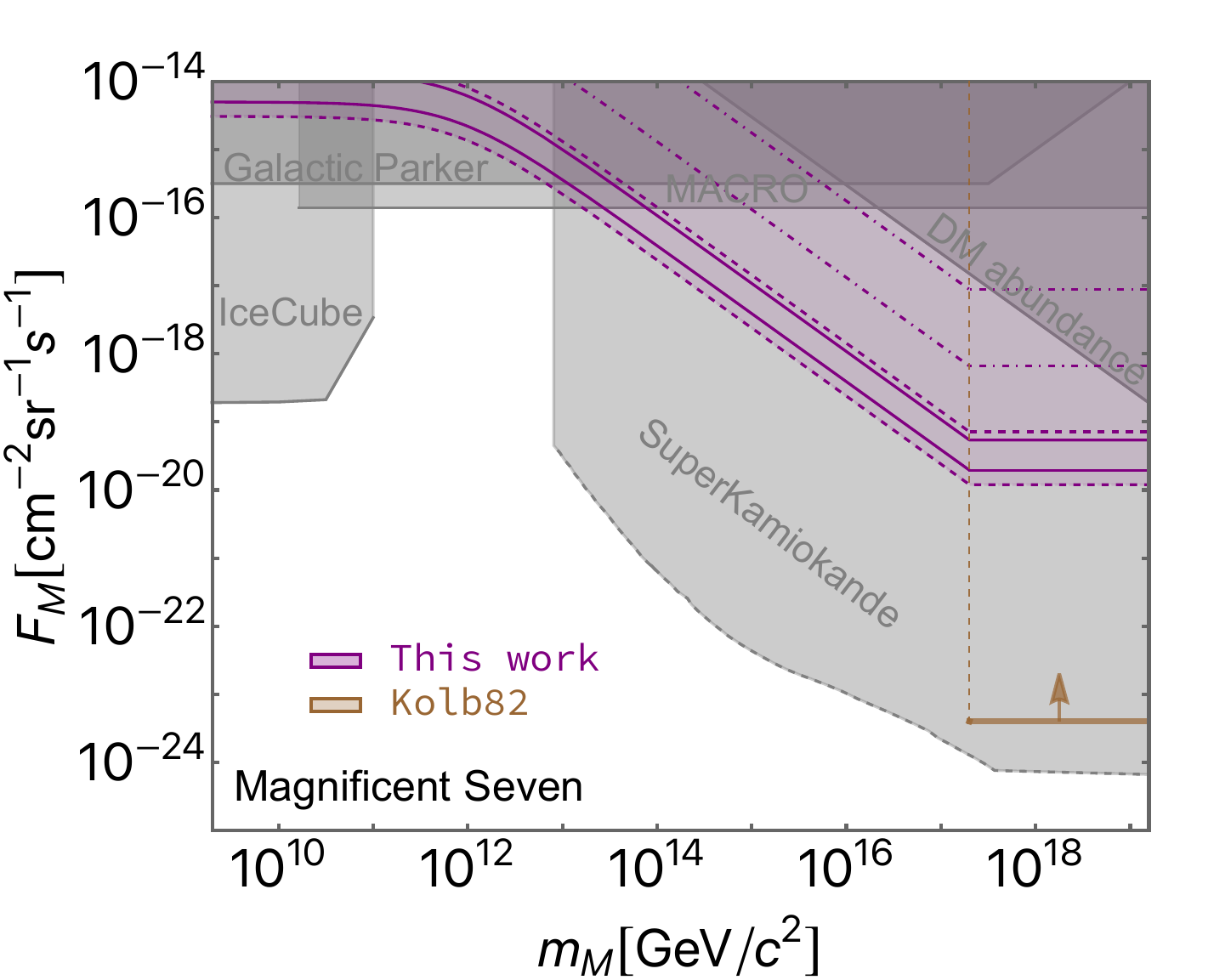}
\caption{\label{fig:result}Comparison of bounds obtained in this work from nucleon decay catalysis using observed X-ray measurements of old isolated millisecond pulsars (\emph{left}), including both measured X-ray fluxes and $3\sigma$ X-ray flux upper limits, and observed redshifted bolometric thermal luminosities of the Magnificent Seven (M7) neutron stars (\emph{right}), with previous bounds in the literature. The solid, dashed, dot-dashed, and dotted lines represent the luminosity distance bins. The parameter space above the lines is excluded, showing that our results provide the most stringent bounds to date on $F_M$ for $m_M$ between $10^{11} - 10^{13}\ {\rm GeV/c^2}$ and are competitive with existing limits.
}
\end{figure*}

GUT monopoles can catalyze nucleon decay with cross-sections that can be comparable to hadronic scales, ca.\ $10^{-27}\ {\rm cm^2}$~\cite{1981ZhPmR..33..658R,Dawson:1982sc,1983NuPhB.212..391C}. While proton decay is a prediction of many GUTs, protons have a lifetime larger than the age of the universe due to the suppression in the decay rate proportional to $M_X^{-4}$, where $M_X$ is the mass of the heavy GUT boson. However, in the presence of GUT monopoles, the decay rate can be enhanced leading to processes like $p + M \rightarrow e^+ + \pi^0 + M$. This mechanism, known as monopole catalyzed nucleon decay or the Callan–Rubakov effect, provides a distinctive signature of GUT monopoles through the production of energetic leptons and mesons. In nucleon-rich environments, the cumulative effects of these decay products can be used to constrain the corresponding monopole flux.

Neutron stars (NSs), with their extremely dense interiors, provide an ideal astrophysical setting for a nucleon rich environment to search for the effects of monopole catalyzed nucleon decay. Monopoles passing through a NS can be captured by the star and subsequently catalyze the decay of a large number of surrounding nucleons. The resulting decay products can then thermalize inside the star, providing an additional heating channel. In steady state, this heating is balanced by thermal photon emission from the stellar surface, leading to a rise in the effective surface temperature.

The resulting thermal radiation makes old NSs calorimetric probes of the monopole flux, providing a direct observational handle~\cite{Kolb:1982si,Harvey:1982py,Dimopoulos:1982cz,Harvey:1982py,Harvey:1985nv,Freese:1983hz}. For sufficiently large monopole-induced heating, the blackbody-like emission produces an observable flux in the soft X-ray band, whereas in the absence of detected X-ray emission, the predicted flux must remain below the corresponding X-ray upper limits (U.L.s). In this \emph{Letter}, for the first time, we apply this idea to actual X-ray observations of old isolated millisecond pulsars (MSPs), using both X-ray measurements and U.L.s from Chandra~\cite{Weisskopf:2001uu}, XMM-Newton~\cite{2001A&A...365L...1J}, and Swift-XRT~\cite{SWIFT:2005ngz} to constrain the monopole flux. By leveraging this method, we manage to put the most stringent constraints (see Fig.~\ref{fig:result}) on certain regions of the parameter space ($F_M - m_M$, where $F_M$ is the monopole flux and $m_M$ is the mass of the monopole).

\textit{\textbf{Capture rate and X-ray flux.}} We begin our analysis by estimating the number of monopoles of a given mass $m_M$ that an old-isolated MSP with mass $M_*$ and radius $R_*$ will accumulate through capture during its lifetime $\tau$ (characteristic age). To do this, we first estimate the velocity of the monopole $v_M$.  Galactic magnetic fields accelerate the monopoles, which typically gain kinetic energy up to $2 \times 10^{11}\ {\rm GeV}$ (see for instance~\cite{Kobayashi:2023ryr}). The Lorentz factor associated to this boost $\gamma_G = 1 + E_M^{\rm kin}/(m_M c^2) \sim 1 + \big(2\times 10^{11}\ {\rm GeV}/(m_M c^2) \big)$ can be used to define the resulting velocity as $v_G(m_M) = c \sqrt{1 - \gamma_G^{-2}(m_M)}$. 
Therefore, the velocity of the monopoles in the Milky Way can be estimated as 
\begin{equation}
\label{eq:velocity}
v_M(m_M) = \max \left[ v_G(m_M), v_{\rm vir} \right] ,
\end{equation}
where $v_{\rm vir} \sim 10^{-3}c$ is the virial velocity of gravitationally bounded objects in the Milky Way. The lower limit on the velocity provided by $v_{\rm vir}$ follows from the fact that the heavier mass monopoles are bounded within the gravitational potential of the Milky Way.

We simplify the analysis by only considering the geometric capture of the monopoles on to the NS. The capture radius $R_{\rm cap}$ is therefore defined as the minimum radial distance from the center of the star that the monopole can pass without getting captured. This can be estimated in the post-Newtonian limit as
\begin{equation}
\label{eq:rcap}
R_{\rm cap}(m_M) = R_* \left( \frac{1 + 2 G M_*/\big(v_M^2(m_M) R_*\big)}{1-R_s/R_*} \right)^{1/2}\,,
\end{equation}
where the Schwarzschild radius $R_s = 2 G M_*/ c^2$. Finally, the number of monopoles that are captured can be defined as $N_M = (2\pi/3) F_M A_{\rm cap} \tau$, where the monopole flux is given by $F_M$ and the effective geometric capture area of the NS as $A_{\rm cap} = \pi R_{\rm cap}^2$. The total luminosity as a result of monopole catalyzed nucleon decay can then be estimated as
\begin{widetext}
\begin{equation}
\label{eq:mono_lum}
\begin{aligned}
L_{\rm cat} &= N_M \left(m_N n_N |v_{\rm MN}|c^2\right)\sigma_{\Delta B}\,
\approx 1.4 \times 10^{36}~\mathrm{erg\ s^{-1}} \\
&\times \Bigg( \frac{F_M}{10^{-18}\ {\rm cm^{-2}s^{-1} sr^{-1}}} \Bigg)
\Bigg( \frac{R_{\rm cap}}{8.7 \times 10^8\ {\rm cm}} \Bigg)^2
\Bigg( \frac{\sigma_{\Delta B}}{10^{-27}\ {\rm cm^{2}}} \Bigg)
\Bigg( \frac{\tau}{10^{10}\ {\rm yr}} \Bigg)
\Bigg( \frac{n_N}{2 \times 10^{38}\ {\rm cm^{-3}}} \Bigg)
\Bigg( \frac{|v_{\rm MN}|}{0.1c} \Bigg)\,,
\end{aligned}
\end{equation}
\end{widetext}
where $m_N$ and $n_N$ are the mass and number density of the nucleons, the relative velocity between the monopoles and the nucleons is given by $v_{\rm MN}$, which we assume as $0.1 c$~\cite{Haensel:2007} similar to the Fermi velocity of the nucleons, and $\sigma_{\Delta B}$ is defined as the cross-section of the monopole-catalyzed nucleon decay process. For the fiducial values of the parameters, $M_* = 1.4 M_\odot$, $R_* = 12$ km, $v_M = 10^{-3} c$, the capture radius is $R_{\rm cap}\sim 8.7 \times 10^8\ {\rm cm}$. Also note that the term in parenthesis $\left(m_N n_N |v_{\rm MN}|c^2\right)$ gives the specific luminosity of the monopole catalyzed decay. The fiducial value of the cross-section is $\sigma_{\Delta B} = \sigma_{\rm QCD} \sim 10^{-27}~\mathrm{cm^{2}}$.

While Eq.~\eqref{eq:mono_lum} represents the total produced luminosity, in this work we are interested only in the fraction of this luminosity emitted in the $0.2-12$ keV X-ray band\footnote{This fraction depends sensitively on the surface temperature of the neutron star, particularly when the lower edge of the instrumental band lies in the Wien tail of the thermal spectrum.}. The choice of the energy range is motivated by its overlap with the sensitivity of several current and forthcoming X-ray facilities. Moreover, NS surface temperatures are expected to range from $10^5~\mathrm{K}$ to $10^7~\mathrm{K}$, depending on their age and thermal evolution, making this spectral band particularly relevant for the present analysis.
For the NSs considered in this work, such temperatures are expected either from standard NS cooling \cite{Ofengeim:2017cum} (for relatively young NSs with ages of $\sim 10^6~\mathrm{yr}$) or from additional late-time heating mechanisms, such as rotochemical heating \cite{Fernandez:2005cg} or heating induced directly by monopole catalysis (for older NSs with ages of $\sim 10^9$–$10^{10}~\mathrm{yr}$).

In order to compare the results from the observations in the X-ray band to the luminosity from monopole-induced catalysis, we compute the fraction of the bolometric luminosity that falls within the X-ray band assuming that monopole-induced heating dominates the thermal luminosity. Assuming a blackbody distribution from the NS surface and accounting for the gravitational redshift, the observed luminosity produced by catalysis in the energy band $(E_1,E_2)$ is
\begin{equation}
\label{eq:lcat_inXray}
L_{\rm cat, X}^{\infty} = \frac{L_{\rm cat}}{(1+z)^2} \times \frac{15}{\pi^4} \int^{E_2/(k_{\rm B} T^{\infty}_{\rm cat})}_{E_1/(k_{\rm B} T^{\infty}_{\rm cat})} \frac{x^3}{e^{x}-1} dx ,
\end{equation}
where $(T^{\infty}_{\rm cat})^4 = L_{\rm cat} / \big(4 \pi R_{*}^2 \sigma_{\rm SB} (1+z)^4 \big)$, $\sigma_{\rm SB}$ is the Stefan-Boltzmann constant, and $z$ is the gravitational redshift defined through $(1+z) = \big(1 - 2GM_*/(R_*c^2)\big)^{-1/2}$, which corresponds to $z \simeq 0.24$ for the fiducial values of $M_*$ and $R_*$ adopted in this work. The corresponding X-ray flux at Earth is $F_{\rm X} = L^{\infty}_{\rm cat, X}/(4 \pi d_{\rm L}^2)$, where $d_{\rm L}$ denotes the luminosity distance to the NS. For sources with measured X-ray fluxes we obtain our bound on the monopole flux from $F_{\rm X} \lesssim F_{\rm X}^{\rm obs}$, whereas for sources with only U.L.s on the X-ray flux we require $F_{\rm X} \lesssim F_{\rm X}^{\rm 3\sigma, lim}$, where $F_{\rm X}^{\rm 3\sigma, lim}$ denotes the derived $3 \sigma$ U.L. on the X-ray flux. For sources with measured bolometric thermal luminosities, such as the M7, we instead directly compare the total luminosity at infinity produced by monopole catalysis with the observed redshifted bolometric luminosity (see Appendix~\ref{appsec:mag7}).

In principle, the decay-induced luminosity can be carried by both photons and neutrinos. In this work, we assume that the energy deposited by the decay products thermalizes efficiently within the neutron-star interior and is subsequently radiated as thermal surface emission. We therefore neglect any direct neutrino energy losses from the decay process and consider only the photon component. This approximation is justified by the fact that the old MSPs considered here are well into the photon-cooling regime, where standard thermal neutrino emission is subdominant to surface photon emission~\cite{Yakovlev:2004iq,Potekhin:2015qsa}.
\begin{table*}
\centering
\caption{List of all old isolated MSPs used for putting constraints in this work (see Fig.~\ref{fig:result}). The table is ordered by increasing $d_L^2/\tau$, where $d_L$ is the luminosity distance and the age $\tau = \min[\tau_c,\tau_{\rm MW}]$. The $3\sigma$ U.L. on the X-ray flux ($F_X^{3\sigma,\rm lim}$) along with the corresponding exposure times ($t_{\rm exp}$) or the measured X-ray flux ($F_X^{\rm obs}$) are listed in the last columns. The old MSPs with significant detection instead of an U.L. are marked with a ``$*$''.}
\label{tab:isolated_old_msps}
\resizebox{\textwidth}{!}{%
\begin{tabular}{c|c|c|c|c|c|c|c|c|c}
\hline\hline
Source &
$d_L$ &
$P$ &
$\tau$ &
$B_{\rm surf}$ &
$d_L^2/\tau$ &
\multicolumn{4}{c}{$F_X^{3\sigma,\rm lim}$ [$10^{-12}\ {\rm erg}\ {\rm cm}^{-2}\ {\rm s}^{-1}$] ($t_{\rm exp}\,[{\rm s}]$)} \\
&
[kpc] &
[ms] &
[$10^{10}$ yrs] &
[$10^8$ G] &
[kpc$^2$ $(10^{10}\,{\rm yr})^{-1}$] &
Chandra Pointed &
Swift XRT &
XMM Pointed &
XMM Slew \\
\hline
J0711-6830 & 0.106 & 5.49 & 0.58 & 2.90 & 0.019 & -- & 0.0576 (3643.1) & $0.0039^{+0.0043}_{-0.0043}$ & 1.043 (10.4) \\
J1709-0333 & 0.214 & 3.52 & 1.40 & 0.82 & 0.033 & -- & 0.1277 (1840.4) & -- & 1.950 (5.90) \\
J0030+0451$^{*}$ & 0.329 & 4.86 & 0.76 & 2.25 & 0.142 & -- & -- & $0.276^{+0.0038}_{-0.0038}$ & 11.46 (1.16) \\
J1744-1134$^{*}$ & 0.395 & 4.08 & 0.72 & 1.93 & 0.217 & $0.01944^{+0.00130}_{-0.00131}$ & 0.1456 (1792.7) & -- & 0.3509 (30.9) \\
J2124-3358$^{*}$ & 0.410 & 4.93 & 0.38 & 3.22 & 0.442 & $0.07858^{+0.00211}_{-0.00225}$ & -- & $0.114^{+0.0048}_{-0.0048}$ & 1.584 (9.51) \\
J1122-3546 & 0.668 & 7.84 & 0.82 & 3.50 & 0.544 & -- & 0.1454 (1737.6) & -- & 1.166 (9.62) \\
J1646-2142 & 0.965 & 5.85 & 1.12 & 2.23 & 0.831 & -- & 0.2693 (604.26) & -- & 1.852 (8.27) \\
J2322+2057 & 0.833 & 4.81 & 0.79 & 2.18 & 0.878 & -- & 0.1231 (1356.4) & -- & 1.204 (9.28) \\
J1710+4923 & 0.506 & 3.22 & 0.28 & 2.45 & 0.914 & -- & 0.0336 (13354) & -- & 1.110 (9.70) \\
J1905+0400 & 1.062 & 3.78 & 1.22 & 1.38 & 0.924 & -- & -- & -- & 1.871 (5.89) \\
J2010-1323 & 1.162 & 5.22 & 1.40 & 1.61 & 0.965 & -- & 0.0653 (3841.2) & -- & -- \\
J1801-1417 & 1.105 & 3.62 & 1.08 & 1.40 & 1.131 & -- & -- & -- & 0.9845 (11.0) \\
J1207-5050 & 1.267 & 4.84 & 1.27 & 1.73 & 1.264 & -- & -- & -- & 0.4414 (33.1) \\
J1629-6902 & 1.111 & 6.00 & 0.95 & 2.48 & 1.299 & -- & -- & -- & 0.9971 (10.9) \\
J0447+2447 & 0.925 & 3.00 & 0.59 & 1.57 & 1.450 & -- & 0.1167 (1913.1) & -- & 0.9589 (11.3) \\
J1721-2457 & 1.392 & 3.50 & 1.00 & 1.41 & 1.938 & -- & -- & -- & 4.426 (8.04) \\
J1103-5403 & 1.683 & 3.39 & 1.40 & 1.13 & 2.023 & 0.00436 & 0.0497 (4021.7) & -- & 1.130 (9.83) \\
J1453+1902 & 1.269 & 5.79 & 0.79 & 2.63 & 2.038 & -- & 0.0223 (9773.4) & -- & 2.268 (6.61) \\
J0154+1833 & 1.622 & 2.36 & 1.28 & 0.84 & 2.055 & -- & -- & -- & 0.995 (11.0) \\
\hline\hline
\end{tabular}
}
\end{table*}

\textit{\textbf{Old MSP catalog \& observational X-ray limits.}} Having established that monopole-catalyzed nucleon decay can produce observable emission in the X-ray band, we now describe the neutron-star samples and X-ray data used to place limits on the monopole flux. 
Our analysis has two complementary components. First, we identify old isolated MSPs (spin period $P \leq 30\ {\rm ms}$)~\cite{Lorimer:2008} that have been observed in X-rays with or without significant X-ray detection. The latter case is considered by using the corresponding $3\sigma$ flux U.L. to place conservative constraints. MSPs are particularly interesting for placing constraints since they typically consist of old recycled (spun up by accretion)~\cite{Bhattacharya:1991} NSs with $\tau \sim 10^9 - 10^{10}\ {\rm yrs}$. The old age is advantageous because the monopole-induced luminosity scales with the accumulated monopole population, $L_M \propto \tau$, while the restriction to isolated systems helps reduce contamination from accretion or binary-related X-ray emission~\cite{Done:2007nc,Becker2009}. Second, as a complementary benchmark, we consider the M7, nearby isolated NSs with direct soft X-ray thermal detections and known surface temperatures~\cite{Haberl:2003cg,Haberl:2006xe,Ertan:2014pfa}. These sources can be analyzed within the same framework, although their measured thermal emission and younger ages lead to weaker monopole-flux limits than those obtained from the old isolated MSP sample. Therefore, in this work, we focus primarily on old-isolated MSPs. The details of the old-isolated MSP selection and the procedure used to obtain the X-ray U.L.s are described below, while the corresponding discussion for the M7 is presented in Appendix~\ref{appsec:mag7}\footnote{Limits on monopole-catalyzed nucleon decay based on the observed X-ray flux of the PSR 1929+10 pulsar were previously proposed in \cite{Freese:1983hz}. However, that analysis was affected by significant uncertainties in the measured properties of the pulsar, which have been later updated~\cite{psrb192910}.}.

The Australian Telescope National Facility (ATNF) Pulsar Catalogue~\cite{Manchester:2004bp,ATNFCatalog} provides a comprehensive database of published pulsars and their measured and derived properties. Having selected the old isolated MSPs from this catalog, we clean the resulting sample by removing MSPs that are in binary systems or globular clusters, are extragalactic, or have existing X-ray associations with an alternative astrophysical source. Finally, we apply a source-quality cut based on the approximate scaling of the expected monopole-flux limit with distance and age. Since, for a fixed X-ray flux, the inferred monopole-flux limit scales roughly as $F_M\propto d_L^2/\tau$, older and nearer MSPs yield stronger constraints. Therefore, we require $d_L^2/\tau < 2.5\ \big(\tilde{d}_L^2/\tilde{\tau}\big)$, where $\tilde{d}_L = 1\ {\rm kpc}$ and $\tilde{\tau} = 10^{10}\ {\rm yrs}$, the same values that were used to obtain the existing monopole-catalyzed nucleon decay bound from Ref.~\cite{Kolb:1982si}. This requirement selects sources that are expected to give limits comparable to, or stronger than, the existing bound\footnote{The numerical factor of $2.5$ is used as a conservative tolerance to retain sources close to the current bound, while removing objects that are unlikely to improve the constraint.}. Therefore, the procedure yields a clean sample of old isolated MSPs with minimal expected X-ray contamination. Note that the age of the old MSP is given by $\tau = \min[\tau_c,\tau_{\rm MW}]$, where $\tau_c$ is the characteristic age and $\tau_{\rm MW} = 1.4 \times 10^{10}\ {\rm yrs}$ is the age of the Milky Way.

To obtain the X-ray U.L.s at the $3\sigma$ level or, when available, the detected X-ray flux, we use archival observations from the most relevant X-ray facilities: XMM-Newton, including both pointed (5XMM-DR15 catalogue)~\cite{Zolotukhin:2016xtf,Webb:2026,XMMSSC} and Slew observations\footnote{Slew observations are taken while the telescope is being slewed from one target to another, rather than during a dedicated pointed observation of a source or sky region. Therefore, they typically have shorter effective exposure times and lower sensitivity than pointed observations.} in the $0.2$–$12~{\rm keV}$ band~\cite{HILIGT}; Swift-XRT in the $0.2$–$12~{\rm keV}$ band~\cite{HILIGT}; and pointed Chandra observations\footnote{For Chandra, we use the CSC (Chandra Source Catalog) broad-band true limiting sensitivity, corresponding to the sensitivity threshold for a source to be classified as a reliable CSC detection rather than a marginal candidate.} in the $0.5$–$7~{\rm keV}$ broad band~\cite{ChandraCSC}. 
When multiple U.L.s or observations are available for a given source, we adopt the one that yields the smallest flux, corresponding to the most sensitive constraint.

Three sources (J1744-1134, J2124-3358, and J0030+0451) have significant X-ray detections in the available archival data. For these objects, the procedure used to derive the constraints on the monopole flux is the same as that adopted for the M7 sample and is described in Appendix~\ref{appsec:mag7}. 
Moreover, we exclude J2031-1254 from the analysis, since no X-ray U.L. could be found for this source. This leaves us with a final sample of $19$ old-isolated MSPs, which is presented in Table~\ref{tab:isolated_old_msps}. Using the neutron-star parameters listed in the table, we then estimate the expected monopole-catalyzed nucleon-decay luminosity for each source.

\textit{\textbf{Results \& constraints.}} The main results of this work are summarized in Fig.~\ref{fig:result}. The gray shaded regions indicate previous constraints on the monopole flux as a function of the monopole mass, from astrophysical and cosmological considerations (the Parker bound \cite{Parker:1970xv,Turner:1982ag} and the dark matter abundance constraint \cite{Planck:2018vyg}) as well as from experimental searches (
IceCube \cite{IceCube:2021eye} and MACRO \cite{MACRO:2002jdv}) for the mass range relevant to this analysis. The mass dependence of the experimental limits originally expressed only in terms of the monopole velocity is taken from \cite{Perri:2025qpg}. We also show, with a dashed contour, the Super-Kamiokande (SK) limit based on the non-observation of a larger solar neutrino flux due to nucleon decay catalysis inside the Sun~\cite{Super-Kamiokande:2012tld}, which are based on the same assumption of this analysis. The SK results are computed as a function of the monopole velocity. Here we rewrite them in terms of the monopole mass using Eq.~\eqref{eq:velocity}.

The \emph{left panel} shows, in blue, the U.L.s on the monopole flux derived from X-ray observations of the selected old MSPs. Constraints obtained from the nearest objects ($d_{\rm L} \leq 0.5~\mathrm{kpc}$) are represented by solid curves, while those from more distant stars are shown with, in order of increasing distance, dashed ($0.5~\mathrm{kpc} <d_{\rm L} \leq 1.0~\mathrm{kpc}$), dot-dashed ($1.0~\mathrm{kpc} < d_{\rm L} \leq 1.3~\mathrm{kpc}$), and dotted curves ($d_{\rm L} > 1.3~\mathrm{kpc}$). The most stringent constraint is provided by J0711-6830, which is the closest NS in our sample. The region above this limit, shaded in blue, is therefore excluded by our analysis. Assuming a catalysis cross section of the order of QCD interactions ($\sigma_{\rm QCD} \sim 10^{-27}\ {\rm cm}^{2}$), these bounds constitute the strongest currently available limits on the monopole flux in the mass range $10^{11}-10^{13}~\mathrm{GeV/c^2}$.
Such a limit can be approximately expressed as
\begin{align}
\label{eq:fm_lim_oldMSP}
F_{\rm M}(m_M)&\left(\frac{\sigma_{\Delta \rm B}}{10^{-27}\ {\rm cm}^{2}}\right) \lesssim  6 \times 10^{-19}~\mathrm{cm^{-2}s^{-1}sr^{-1}} \\ 
& \times \max \left[4 \times 10^{-6}, \min\left[ \frac{2 \times 10^{11}~\mathrm{GeV/c^2}}{m_{\rm M}} , 1 \right] \right] . \nonumber 
\end{align}

The \emph{right panel} shows, in purple, the U.L.s on the monopole flux derived from the observed X-ray emission of the six selected candidates of the M7 NSs. The line styles follow the same distance-based convention adopted for the old MSP sample (solid: $d_{\rm L} \leq 0.25~\mathrm{kpc}$, dashed: $0.25~\mathrm{kpc} < d_{\rm L} \leq 0.35~\mathrm{kpc}$, dot-dashed: $d_{\rm L} > 0.35~\mathrm{kpc}$). Although these nearby isolated NSs have measured thermal X-ray emission, their younger ages and larger observed thermal luminosities make the resulting constraints weaker than those obtained from the old isolated MSP sample (Table~\ref{tab:isolated_old_msps}). The strongest limit is obtained from RX J0420.0-5022, and the region above this curve, shaded in purple, is excluded by the results of the analysis.

Finally, for comparison, we include in the plots the result of Ref.~\cite{Kolb:1982si}, shown as a brown segment (``Kolb82'' in the legend). In their analysis, the authors assumed a monopole velocity $v_{\rm vir}$. Hence, with this choice, their original results are valid for the results of this work only in the limit of very large monopole masses, $m_M\gtrsim 10^{17}~\mathrm{GeV/c^2}$. Furthermore, unlike in our analysis, the authors do not account for the fact that only a fraction of the photons produced by catalysis are emitted in the X-ray band. This fraction can be significantly reduced depending on the stellar temperature which can substantially weaken the inferred constraint.

The bounds shown in Fig.~\ref{fig:result} exhibit three distinct mass dependences, which can be understood as follows. For small monopole masses, $m_{\rm M} \ll 2 \times 10^{11}~\mathrm{GeV}/c^2$, monopoles move at relativistic velocities, $v_{\rm M} \sim v_{\rm G} \sim 1$. In this regime, the capture radius is independent of the monopole mass, and consequently the flux limit is also mass independent. For intermediate masses, where $v_{\rm G}$ becomes non-relativistic but still satisfies $v_{\rm G} > v_{\rm vir}$, the monopole velocity scales as $v_{\rm M} \propto m_{\rm M}^{-1/2}$. The capture radius then scales as $R_{\rm cap} \propto m_{\rm M}^{1/2}$, implying a flux bound $F_{\rm M} \propto m_{\rm M}^{-1}$. As a result, the constraint becomes increasingly stringent with increasing monopole mass. Finally, for extremely heavy monopoles, such that $v_{\rm G} < v_{\rm vir}$, the velocity saturates at $v_{\rm vir}$ and becomes independent of the monopole mass. Therefore, $R_{\rm cap}$ is also independent of the monopole mass, and the resulting flux limit approaches again a constant value.

For Fig.~\ref{fig:result}, we only show $m_M \gtrsim 10^{10}~\mathrm{GeV/c^2}$. This choice reflects the fact that only monopoles produced during the breaking of a GUT symmetry are expected to mediate nucleon decay catalysis at leading order through excited states of GUT gauge vector bosons within the monopole core. Nevertheless, monopoles with lower masses, arising in scenarios with two-step GUT symmetry breaking, have also been argued to induce nucleon decay catalysis through alternative mechanisms, such as higgs or higgsino exchange, as well as higher-dimension operators \cite{Kephart:2001ix}. Our choice of mass range is therefore conservative and could be extended for models predicting intermediate or low-mass monopoles that are capable of catalyzing nucleon decay.

The analysis presented in this work assumes that monopole–antimonopole annihilation inside NSs is always negligible. The conditions under which this assumption is valid were studied in \cite{Harvey:1982py}, where it was shown that annihilation can be neglected for a typical NS. We repeated the analysis for the NSs considered in this work and confirmed that monopole–antimonopole annihilation is negligible for the purposes of this work. The details of our calculation are presented in Appendix~\ref{appsec:pair_ann_ns}.

\textit{\textbf{Conclusions \& outlook.}} In this work, for the first time, observational X-ray data from old isolated MSPs are used to derive constraints on the Galactic flux of magnetic monopoles catalyzing nucleon decay. We use archival data from XMM-Newton, Chandra, and Swift-XRT (both significant observations and $3\sigma$ flux U.L.s), to place stringent constraints on the monopole flux in the $F_M$--$m_M$ plane for a given monopole-catalysis cross section. Besides implementing a more accurate methodology, our limits improve on the previous bound from Ref.~\cite{Kolb:1982si}, and extend the results to smaller masses. Our limits are also competitive with the constraints from Super-Kamiokande, based on the same assumptions for the nucleon decay catalysis. In particular, in the case of a cross section comparable to QCD processes, NSs provide the strongest limits to date in the mass range $10^{11}-10^{13}~\mathrm{GeV/c^2}$ (see Fig.~\ref{fig:result}). This constraining power follows from two key properties of NSs: their nucleon-rich dense interiors and their ability to efficiently capture monopoles over their lifetimes. 

%
\begin{figure}
\centering
\includegraphics[width=0.49\textwidth]{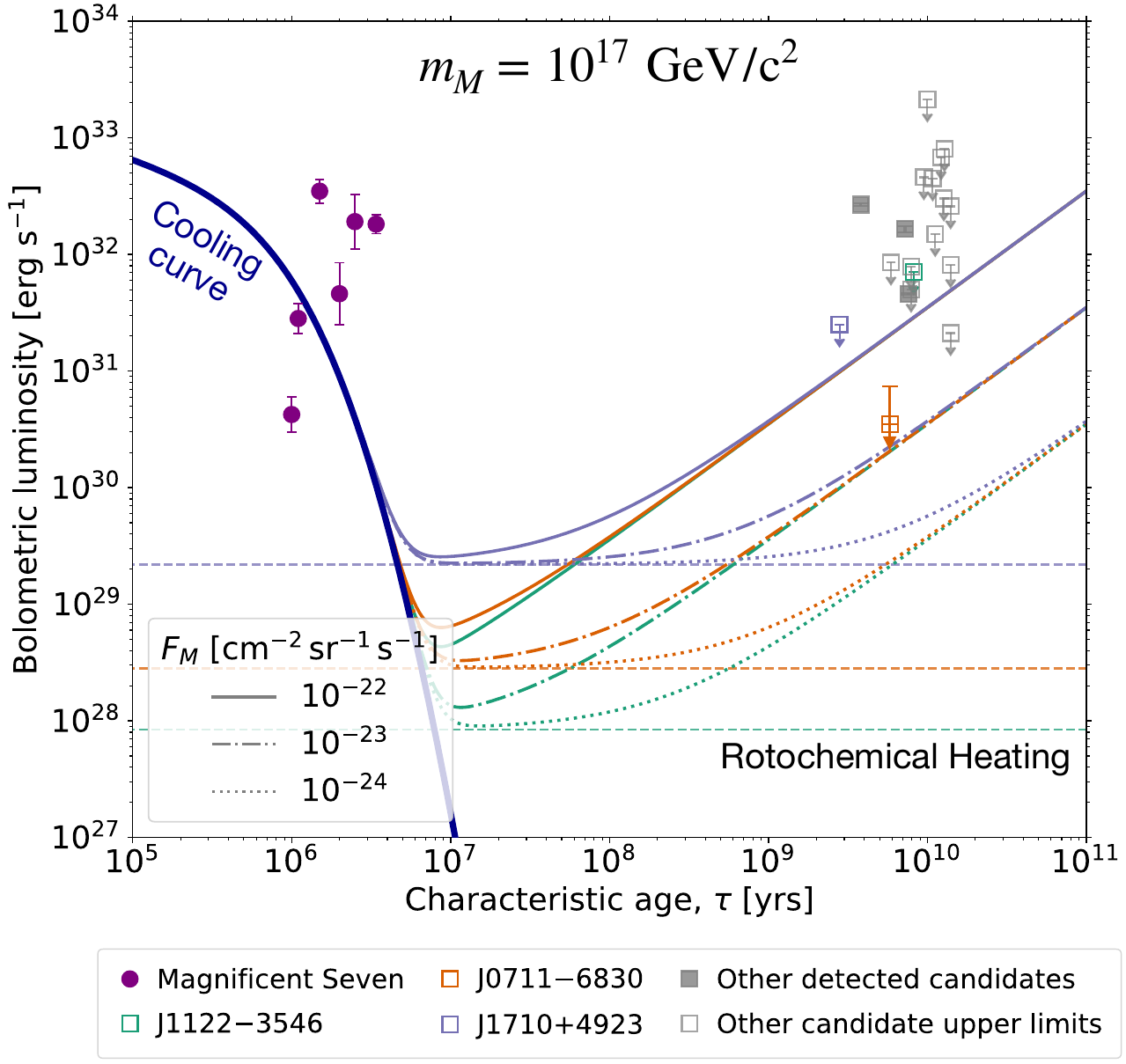}
\caption{\label{fig:heating}Standard cooling curve (solid dark-blue), rotochemical-heating plateaus (dashed colored lines), and the corresponding total bolometric luminosities including monopole-induced heating (colored curves) for representative monopole fluxes $F_M$. The rotochemical-heating and total-luminosity curves are shown for clarity only for three representative MSPs spanning the range of rotochemical heating levels in our sample. The observed bolometric thermal luminosities of the detected MSPs are shown as filled squares, the $3\sigma$ U.L.s as open squares, and the M7 as purple filled circles.
}
\end{figure}

Since the monopole-induced luminosity grows with the accumulated monopoles, the constraining power improves for older and nearer NSs with deeper X-ray observations. Dedicated long-exposure pointed observations of promising old isolated MSPs could therefore significantly improve these limits in the future. In particular, observations of the MSPs considered in this work with the sensitive instruments of XMM-Newton and Chandra (especially J1709-0333, which has the lowest value of the ratio $d_{\rm L}^2/\tau$ among the stars for which neither observatory has yet provided an U.L. on the X-ray flux) could improve the current constraint by up to two orders of magnitude, possibly surpassing the sensitivity of Super-Kamiokande even in the high-mass regime.

If X-ray counterparts to more of these MSPs are detected through dedicated searches and the MSP cooling histories are sufficiently well constrained, a deviation from the expected luminosity–age relation could provide evidence for an additional heating mechanism, with monopole-induced heating being one possible explanation.
In Fig.~\ref{fig:heating}, we illustrate how targeted X-ray observations of old NSs could reveal a characteristic signature of monopole-induced heating. 

At early times, the thermal evolution is dominated by neutrino emission, while photon emission from the stellar surface becomes the primary cooling channel at later epochs~\cite{Ofengeim:2017cum}. The figure shows in dark-blue the representative standard-cooling curve based on Eq.~(15) of Ref.~\cite{Ofengeim:2017cum}. The measured bolometric luminosities of the M7 are broadly consistent with this behavior, and the cooling curve exhibits a rapid decline for ages $\tau\gtrsim10^{6}\ {\rm yrs}$.
At late times, conventional reheating mechanisms can partially compensate for this decline. In particular, rotochemical heating arises because spin-down driven compression perturbs the stellar matter away from beta equilibrium; the subsequent non-equilibrium weak reactions release heat as the composition relaxes back toward equilibrium~\cite{Fernandez:2005cg}. For illustration, we adopt the quasi-equilibrium estimate given in Eq.~(68) of Ref.~\cite{Fernandez:2005cg}, evaluated using the measured $P$ and $\dot P$ values of each MSP, and represent the resulting contribution as a source-dependent luminosity plateau\footnote{In a complete thermal-evolution calculation, the rotochemical luminosity is not strictly constant and typically decreases as the spin-down power declines; see, e.g., Ref.~\cite{Yanagi:2019vrr}.}. We then combine the standard-cooling, rotochemical-heating, and monopole-induced contributions for several representative values of $F_M$.

In contrast to both standard cooling and rotochemical heating, the monopole-induced luminosity increases with the accumulated monopole population, scaling as $L_M\propto\tau$. This leads to a characteristic late-time upturn of the luminosity. Such a qualitatively distinct age dependence provides a direct diagnostic, that is, with sufficiently sensitive observations of old isolated NSs, a monopole-induced heating component could be distinguished from the standard cooling trend and from conventional reheating mechanisms. Since more realistic thermal-evolution models generally refine the expected astrophysical heating contribution on a source-by-source basis, they could further strengthen the monopole-flux limits derived here. Our simplified treatment should therefore be regarded as conservative.

In conclusion, our results demonstrate that old NSs constitute a natural and powerful laboratory for constraining the abundance of magnetic monopoles in the Universe. Future observations could significantly tighten these constraints, opening new avenues for both theoretical investigations and experimental searches aimed at the direct detection of magnetic monopoles. Should an X-ray flux be observed, distinguishing the standard thermal emission from a contribution induced by monopole-catalyzed processes will require a more detailed understanding of the phenomenological impact of such catalysis on NS evolution. Establishing an indirect signature of magnetic monopoles will therefore necessitate further dedicated studies, which we leave for future work.
%

\acknowledgments
\textit{\textbf{Acknowledgements.}}
We thank Karri I. I. Koljonen, Daniele Rogantini, and Irina Zhuravleva for suggestions and useful discussions. M.\,M. acknowledges support from the FermiForward Discovery Group, LLC under Contract No. 89243024CSC000002 with the U.S. Department of Energy, Office of Science, Office of High Energy Physics.
D.P. was partially supported by the National Science Centre, Poland, under research grant no. 2020/38/E/ST2/00243.
\bibstyle{apsrev4-2}
\bibliography{refs}
\clearpage
\newpage
\appendix
\makeatletter
\@removefromreset{equation}{section}
\makeatother
\twocolumngrid
\maketitle
\onecolumngrid
\begin{center}
	\textbf{\Large Supplementary Material}
	 \bigskip\\
		\textbf{\large Constraints on magnetic monopoles from X-ray observations of neutron stars}
		 \medskip\\
   {Mainak Mukhopadhyay, Daniele Perri, and Edward W. Kolb}
\end{center}
\renewcommand{\thesection}{S\arabic{section}}
\renewcommand{\theequation}{S\arabic{equation}}
\renewcommand{\thefigure}{S\arabic{figure}}
\renewcommand{\thetable}{S\arabic{table}}
\renewcommand{\thepage}{S\arabic{page}}
\setcounter{equation}{0}
\setcounter{figure}{0}
\setcounter{table}{0}
\setcounter{page}{1}
\setcounter{secnumdepth}{4}
\section{Magnificent Seven (M7)}
\label{appsec:mag7}
In this appendix, we provide additional details on the Magnificent Seven (M7)~\cite{Haberl:2003cg,Haberl:2006xe,Ertan:2014pfa} and the constraints that can be obtained from their observed X-ray flux. The M7 are a nearby population of radio-quiet isolated neutron stars (NSs) discovered through their soft thermal X-ray emission.
The M7 are especially useful as benchmark calorimeters because their X-ray spectra are dominated by blackbody like thermal emission with temperatures ($k_{\rm B} T^\infty$) between $45 - 100$ eV, and exhibit no significant non-thermal component. In comparison to the MSPs in Table~\ref{tab:isolated_old_msps}, the M7 are relatively young, with an age of approximately $10^{5} - 10^{6}$ years. Since the monopole-induced luminosity scales with the accumulated number of  monopoles, the comparatively smaller ages lead to weaker limits than those obtained in the main text. Nevertheless, because their thermal X-ray fluxes are directly measured and the sources are nearby, they provide an independent and conservative check on the effects of monopole-induced heating.

For these sources, we impose the constraint that the total monopole-induced luminosity measured by an observer at infinity not exceed the observed redshifted bolometric thermal luminosity corresponding to the observed X-ray flux~\cite{Potekhin:2020ttj}, that is, $L_{\rm cat}/(1+z)^2 \leq  L_{\rm bol}^{\infty,\rm obs}$ (see Eq.~\ref{eq:mono_lum} for $L_{\rm cat}$). This is the appropriate comparison because the energy released through monopole-catalyzed nucleon decay is assumed to thermalize in the stellar interior and therefore contribute to the total thermal surface luminosity rather than to a particular
instrumental energy band. We also quote the corresponding bolometric flux $F_{\rm bol}^{\infty,\rm obs} = L_{\rm bol}^{\infty,\rm obs}/(4\pi d_L^2)$ to facilitate comparison with the flux constraints used for the MSP sample.

The M7 sources and their relevant parameters are summarized in Table~\ref{tab:m7_sources}. For all the M7 stars we assume a mass of $M_* = 1.4 M_\odot$, radius of $R_*  = 12\ {\rm km}$, and moment of inertia $I \sim M_* R_*^2 = 4 \times 10^{45}\ {\rm g\ cm^2}$. The spindown rate $\dot{P} = dP/dt$ is used to compute the characteristic age of the NS, $\tau_c = P/\big( 2 \dot{P} \big)$, where $P$ is the spin period. The observed thermal emission is commonly described in terms of a redshifted surface temperature, $T^\infty = T_s \lambda_z$, where $T_s$ is the local surface temperature, $\lambda_z = 1/(1+z) = \sqrt{1- R_S/R_*}$ is the gravitational redshift correction, $R_S = 2 G M_*/c^2$ is the Schwarzschild radius of the NS, and $k_B$ is the Boltzmann constant. For a blackbody fit, the bolometric thermal luminosity inferred by an observer at infinity can be written as $L_{\rm bol}^{\infty} = 4\pi (R^\infty)^2 \sigma_{\rm SB} (T^{\infty})^{4}$, where $R^{\infty}$ is the apparent blackbody radius and $\sigma_{\rm SB} = 5.67 \times 10^{-5}~\mathrm{erg~cm^{-2}s^{-1}K^{-4}}$ is the Stefan-Boltzmann constant. Finally, the surface dipolar magnetic field strength can be computed as
\be
B_{\rm surf} = \left( \frac{3 c^3 I}{8 \pi^2 R_*^6} P \dot{P} \right)^{1/2} \approx 10^{13}\ {\rm G} \left( \frac{I}{4 \times 10^{45}\ {\rm g\ cm}^{2}} \right)^{1/2} \left( \frac{R_*}{12\ {\rm km}} \right)^{-3} \left( \frac{P}{1~\mathrm{s}} \right) \left( \frac{10^{6}~\mathrm{yrs}}{\tau} \right)^{1/2} \,.
\ee
Furthermore, the apparent radius $R^\infty$ measured by a distant observer is related to the physical radius $R_*$ by $R_\infty = R_*/\lambda_z$. Assuming a NS mass of $M_* = 1.4 M_\odot$ corresponds to a physical radius of $R_* \sim 11 - 13$ km. Therefore, our fiducial value of $R_* = 12$ km sets $R^\infty \approx 14.8$ km. 

A few characteristics of the M7 sources listed in Table~\ref{tab:m7_sources} can be noted. 
We exclude RX J1605.3+3249 from our analysis, since it does not have a confirmed spin period. This is because, unlike the other M7 sources, it exhibits only very weak pulsations. Finally, the constraints that we obtain for M7 are presented in Fig.~\ref{fig:result} \emph{(right panel)} and discussed in the main text below Eq.~\eqref{eq:fm_lim_oldMSP}.
Because the M7 constraints are limited by their detected thermal luminosities rather than by instrumental non-detection thresholds, substantially deeper observations would not improve the bounds in the same direct manner as for the undetected MSPs. Improved measurements of their distances and thermal spectra could, nevertheless, reduce the associated astrophysical uncertainties.
\begin{table*}[ht!]
\centering
\caption{List of Magnificent Seven (M7) isolated NSs used as complementary thermal-emission constraints in this work. The spin period $P$, characteristic age $\tau$, and surface magnetic field strength $B_{\rm surf}$ are taken from Ref.~\cite{Bogdanov:2024kjp}. The luminosity distance $d_L$, temperatures $T^\infty$, and bolometric thermal luminosities $L_{\rm bol}^\infty$ are taken from Ref.~\cite{Potekhin:2020ttj}. The fluxes shown in the last column are effective bolometric fluxes, $F_{\rm bol}^{\infty}=L_{\rm bol}^{\infty}/(4\pi d_L^2)$, included only to facilitate comparison with the MSP flux U.L.s in Table~\ref{tab:isolated_old_msps}. We remove RX J1605.3+3249 since it does not have a confirmed spin period.}
\label{tab:m7_sources}
\resizebox{\textwidth}{!}{%
\begin{tabular}{c|c|c|c|c|c|c|c}
\hline\hline
Source &
$d_L$ [kpc] &
$P$ [s] &
$\tau$ [$10^6$ yrs] &
$B_{\rm surf}$ [$10^{13}$ G] &
$k_B T^\infty$ [eV] &
$L_{\rm bol}^{\infty}$ [$10^{30}\ {\rm erg}\ {\rm s}^{-1}$] &
$F_{\rm bol}^{\infty}$ [$10^{-12}\ {\rm erg}\ {\rm cm}^{-2}\ {\rm s}^{-1}$] \\
\hline
RX J0420.0--5022
& $0.325$--$0.345$
& $3.45$
& $1.9$
& $1.0$
& $45.0\pm 2.6$
& $6\pm 2$
& $0.30$--$0.60$ \\

RX J0720.4--3125
& $0.286^{+0.270}_{-0.230}$
& $8.39$
& $1.9$
& $2.5$
& $90$--$100$
& $190^{+130}_{-80}$
& $11.2$--$32.7$ \\

RX J0806.4--4123
& $0.235$--$0.250$
& $11.37$
& $17.0$
& $1.1$
& $90$--$110$
& $16$--$25$
& $2.1$--$3.8$ \\

RX J1308.6+2127
& $0.380^{+0.020}_{-0.030}$
& $10.31$
& $1.5$
& $3.4$
& $50$--$90$
& $330^{+50}_{-70}$
& $15.1$--$22.0$ \\

RX J1856.5--3754
& $0.123^{+0.011}_{-0.015}$
& $7.06$
& $3.7$
& $1.5$
& $36$--$63$
& $50$--$80$
& $27.6$--$44.2$ \\

RX J2143.0+0654
& $0.390$--$0.430$
& $9.43$
& $3.6$
& $2.0$
& $40$--$100$
& $50$--$170$
& $2.5$--$8.5$ \\

\hline\hline
\end{tabular}
}
\end{table*}
\section{Pair annihilation inside the neutron stars}
\label{appsec:pair_ann_ns}
The computation of this work assumes that the effects of monopole-antimonopole annihilation inside the NS are always negligible. This assumption depends on the inner structure of the NSs. In particular, in this appendix we discuss the two cases of a NS with a normal interior, and the one of a superconducting one, demonstrating that in all cases considered any effect of annihilation can be neglected.

In the case of a normal NS interior, in Ref.~\cite{Harvey:1982py} it was shown that monopole annihilation is suppressed by the presence of strong magnetic fields. In particular, minimum interior magnetic field required for this suppression to hold is given by
\begin{equation}
\label{eq:B_cond_ann}
B \gtrsim 3 \times 10^2 N_{\rm M}^{1/3} \left( \frac{m c^2}{10^{16}~\mathrm{GeV}} \right)^{2/3} \mathrm{G} ,
\end{equation}
where as before $N_{\rm M}$ is the number of monopoles accumulated inside the NS. Assuming that two-body annihilation is the dominant process reducing the monopole population, whenever annihilation is relevant the number of monopoles evolves as $N_{\rm M} = N_{\rm M}^{\rm eq} \tanh{(t/t_0)}$ with the equilibrium value given by
\begin{equation}
N_{\rm M}^{\rm eq} \sim 2 \times 10^{17} \left( \frac{F_{\rm M}}{10^{-14}~\rm cm^{-2} s^{-1} sr^{-1}} \right)^{1/2} \left( \frac{k T_{\rm M}}{200~\mathrm{MeV}} \right) .
\end{equation}
Here $T_{\rm M}$ is the effective temperature of the monopole system, assuming it can be treated as an ideal gas, with a typical value of $kT_{\rm M} \sim 200~\mathrm{MeV}$. For the typical interior magnetic fields of NSs ($B \sim 10^{12} - 10^{15}~\mathrm{G}$)~\cite{Braithwaite:2006,Ferrario:2015}, the monopole flux considered in this work ($F_{\rm M} \lesssim 10^{-14}~\mathrm{cm^{-2} s^{-1} sr^{-1}}$), and assuming $N_{\rm M} = N_{\rm M}^{\rm eq}$, the condition in Eq.~\eqref{eq:B_cond_ann} is satisfied for $m_{\rm M} \lesssim 10^{22}~\mathrm{GeV}$.
This range encompasses all the masses considered in Fig.~\ref{fig:result}, confirming the validity of our approximation. For the M7, the assumption $N_{\rm M}=N_{\rm M}^{\rm eq}$ is not necessarily valid, since the accumulated monopole number may not have reached saturation for the age of these NSs. Nevertheless, because $N_{\rm M} \lesssim N_{\rm M}^{\rm eq}$ by construction, using the equilibrium value provides a conservative estimate. This is because a smaller monopole population relaxes the condition in Eq.~\eqref{eq:B_cond_ann}, thereby extending the range of monopole masses for which annihilation can be neglected.

Finally, in the case of a superconductive behavior, in Ref.~\cite{Harvey:1982py} it was shown that the monopoles inside the NS are confined to magnetic field flux tubes. Therefore, as long as the number of flux tubes is larger than the number of monopoles inside the NS, annihilation provides a negligible contribution to the computation of the monopole abundance. The estimated number of flux tubes is of order $10^{31}$, and therefore we can safely assume that any effect to the results of this paper from monopole annihilation is negligible in the case of a superconductive interior.
\end{document}